# Building Drift

## Documenting On-Site Construction Adaptations Across Material Lifecycles

**Ritik Batra**
Cornell Tech &
CITA Royal Danish Academy of Fine Arts

**Martin Tamke**
CITA Royal Danish Academy of Fine Arts

**Tom Svilans**
CITA Royal Danish Academy of Fine Arts

**Jan Hüls**
CITA Royal Danish Academy of Fine Arts

**Amritansh Kwatra**
Independent

**Steven J. Jackson**
Cornell University

**Thijs Roumen**
Cornell Tech

**Mette Ramsgaard Thomsen**
CITA Royal Danish Academy of Fine Arts

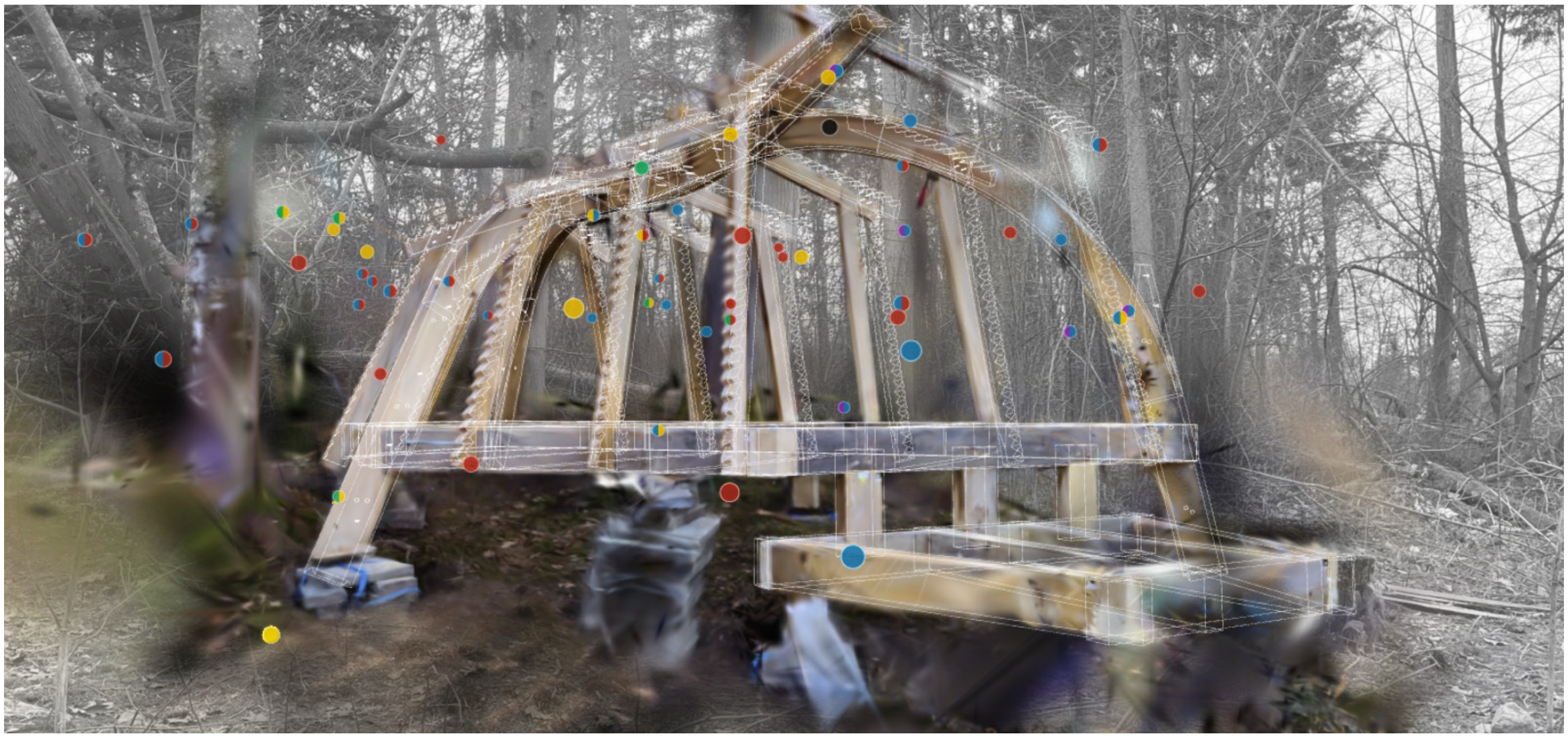

1 Pentimento aligns the *as-built* structure as Gaussian Splats with the *as-designed* model using overlaid on-site adaptations and comments.

### ABSTRACT

In a circular economy for construction, reclaimed materials carry prior lives of use and go on to have post-lives in future buildings. Yet working with such materials introduces unpredictability that requires on-site improvisation, making their reuse challenging to document and scale across building lifetimes. Without documentation, the on-site adaptations that make construction with reclaimed materials possible leave collaborators, evaluators, and inheritors without the information they need to continue, assess, and reuse materials. We define the collective deviation of the physical state from the digital model through these adaptations as "building drift." Through a case study, ReShelter, a reclaimed timber pavilion constructed in the forest, we develop a preliminary description of building drift as emerging from five practices essential to building with reclaimed materials: Tending the Site, Foraging for Fit, Interpreting the Material, Marking Measurements, and Coordinating Across Communities.

We then present Pentimento, a documentation tool that leverages video documentation and 3D Gaussian Splatting to spatially, temporally, and semantically represent on-site adaptations in relation to the designed model. Pentimento enables each stakeholder to navigate material histories in ways that reduce barriers to material reuse. Together, these contributions open pathways towards computational tools that support the on-site improvisation essential to construction with reclaimed materials, enabling more sustainable cycles of recovery, repair, and reuse.

## INTRODUCTION

Imagining a circular economy of construction materials requires the recovery, repair, and reuse of materials across multiple building lifetimes. Yet the accomplished ("as-built") form of buildings, like other forms of human endeavor, are rarely executed exactly to plan (Suchman 1987; Siza 1997). Unlike standardized and pre-processed materials with known properties (Gordon and Calvo 2021), reclaimed materials carry unique heterogeneities such as knots, surface contamination, and embedded nails that render traditional models of tolerance, joinery, and geometry insufficient (Pelin et al. 2024). The gap between the *as-designed* model and *as-built* state, referred to as the *sim-to-real* gap (Peng et al. 2018), therefore exists before a construction project begins: how each material was stored, how it was processed, and what it had become in a previous project. On site, builders make adaptations and improvisations around such heterogeneities, further widening the gap and compounding challenges for material reuse (Golparvar-Fard et al. 2011).

We call the collective deviation of the physical state from the digital model through on-site adaptations and improvisations *building drift*. Grounded in a case study of ReShelter (Tamke et al. 2026), a reclaimed timber pavilion constructed in the forest, we describe five essential sources of building drift: Tending the Site, Foraging for Fit, Interpreting the Material, Marking Measurements, and Coordinating across Communities. To make these sources explicit, we developed Pentimento, a documentation tool that aligns on-site adaptations with the original CAD model (shown in Figure 2) and makes them selectively legible to different viewers. This work sits at the intersection of computational design, human-computer interaction, and science and technology studies, demonstrating how 3D reconstruction, voice transcription, and LLM-powered categorization can be combined to document building drift and therefore reduce barriers to using reclaimed materials in real-world construction projects.

The practices of building drift are typically undocumented, creating critical consequences that propagate across a material's lifetime. Consider a reclaimed timber beam that arrives with a knot where a screw hole was planned. The builders then drill a new hole on site that can complete the connection but avoids the knot. The beam may be successfully installed but its modified state is not recorded. The consequences compound: a collaborator (another builder) drills another hole at the original location, an evaluator (deconstruction crew) scraps the material, and an inheritor (designer) models a connection assuming the original geometry only to discover the mismatch on site, forcing yet another undocumented adaptation. Each of these stakeholders (collaborators, evaluators, and inheritors) needs different information to maintain a sustainable ecosystem of material reuse: collaborators need to know why, evaluators need to know how, and inheritors need to know the final geometry (Browne and Popovic Larsen 2024). Without this record, each stakeholder is left to rediscover what others may already know, reducing working knowledge (Orr 1990; Harper 1987) and quietly eroding the material's reuse potential.

These individual consequences compound further in a circular economy, where building drift propagates beyond a single project's designers, builders, and engineers to encompass a material lifecycle across projects and decades (Thomsen et al. 2026). Materials arrive with *pre-lives* of prior engagements, change and evolve through the course of building lifetimes (Brand 1995), and go on to have *post-lives* in future construction projects, with each transition introducing new stakeholders with different needs and incomplete knowledge (Tamke et al. 2025). Where sustainable construction and fabrication is often imagined as static, or a fixed world that can be copied and pasted across space and time, building drift highlights a world that is dynamic and specific, always in motion and always being re-negotiated.

To integrate our dynamic world with digital models, we need tools for documenting on-site construction work in ways that enable each stakeholder to act on the information relevant to them. Existing approaches struggle with this complex task: capturing the outcome leaves stakeholders without the context they need (Ward and Chapman 2008), while capturing everything produces records too dense to navigate and too burdensome to create on site (Sun et al. 2017). What is needed is a lightweight documentation tool

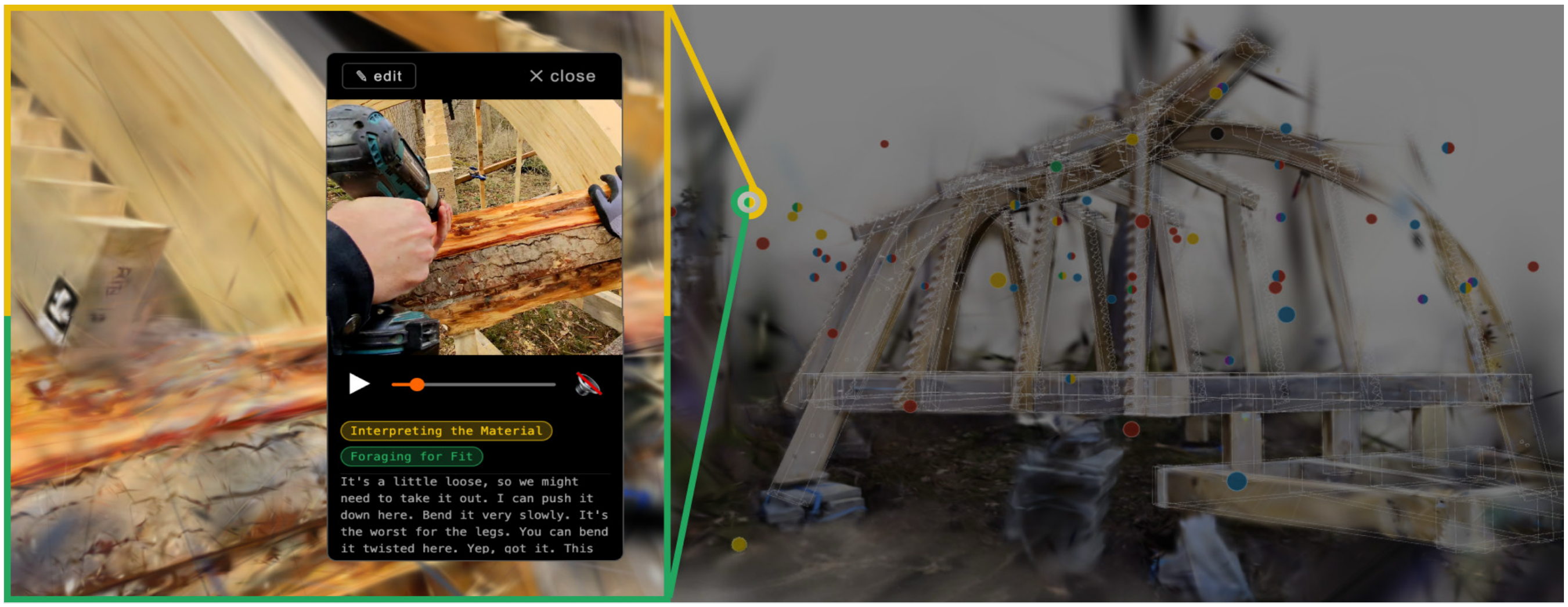

2 Aligned on-site adaptation in Pentimento.

that spatially and temporally grounds on-site adaptations and makes them selectively legible to each stakeholder. With Pentimento, builders can document their adaptations through video and audio recordings captured using a mobile device. Pentimento then processes the videos by generating 3D Gaussian Splats (3DGS) (Kerbl et al. 2023) and aligns them with the original CAD model (Kwatra et al. 2025). Stakeholders separated by space and time can navigate construction project states and adaptations spatially, temporally, and semantically to further support the recovery, repair, and reuse of reclaimed materials across building lifetimes.

Through this work, we make two central contributions. First, we offer examples and a preliminary classification for sources of building drift from a case study of a reclaimed timber construction project. Second, we develop Pentimento, a computational tool that demonstrates how to document sources of building drift by aligning on-site adaptations with the original 3D model. In addition, we call attention to *in-situ* and "as-found" commitments of architecture and building practice that may be missed in "blueprint-first" imaginations of design and that any effective movement towards circularity must find ways to accommodate (De Caigny et al. 2023). Together, these contributions create pathways towards computational tools that preserve and surface material histories across their lifecycles; and modes of architectural and building practice that can support more sustainable cycles of recovery, repair, and reuse.

## STATE OF THE ART

### Computational Models for Reclaimed Materials

Computational models for architecture assume material homogeneity. Popular Computer-Aided Design (CAD) software such as Rhino3D or Autodesk Fusion abstract materials away as uniform, homogeneous surfaces. Abstract models reflect an implicit assumption that, for example, a timber beam is always consistent in geometry, tolerance, and structural behavior. The resulting construction process reflects what Ingold calls a hylomorphic model of making where form is imposed onto passive matter (Ingold 2009). Reclaimed materials, however, break this abstraction: a timber beam carries nails from prior connections, sparse regions of moisture damage, and variable grain that render standard tolerance and joinery models insufficient (Tamke et al. 2024). Unlike standardized materials that arrive with datasheets, reclaimed materials must have their identity reconstructed as the builder *"bind[s] their own pathways or lines of becoming into the texture of material flows comprising the lifeworld"* (Ingold 2009). Heterogeneity is a defining characteristic of reclaimed materials and why they are often discarded (Liboiron and Lepawsky 2022). This positions drift as an inevitable feature of construction with reclaimed materials, as builders negotiate with materials, environments, and each other in ways that computational models cannot anticipate.

### Construction as Collaborative Work

Architecture has long grappled with the gap between as-designed and as-built (Golparvar-Fard et al. 2011; Wang et al. 2025). Brand describes buildings as living processes that evolve through use (Brand 1995), and later Cairns and Jacobs build on that idea by describing architecture's *"delusions of permanence,"* arguing that decay and death are not failures but productive forces that should reshape how we think about building (Cairns and Jacobs 2014). Research at the intersection of architecture and repair studies has emphasized the essential work of maintenance and repair in shaping and reshaping buildings and their materials (Bovet and Ignaz 2019; Jackson et al. 2024; Thomsen et al. 2024; DeSilvey 2017). As this work makes clear, connecting across building lifetimes, from inception to end of life, requires the coordinated efforts of many actors, including designers, builders, engineers, maintainers, managers, and inhabitants, each holding partial knowledge of the whole (Orr 1996). Beyond this human cast, materials warp and degrade, environment terrain shifts, and weather reworks the whole. Building drift is thus an accumulated record of negotiation between human intent and the myriad resistances of the world.

### Documentation for On-Site Adaptations

Existing approaches to documenting construction projects struggle with the complexity of drift. Computer vision-based approaches assume controlled environments and generally struggle to handle the heterogeneous materials and unpredictable conditions of real construction sites (Teizer 2015). Extended Reality (XR) guides builders by overlaying the digital model onto the physical site (Alizadehsalehi et al. 2020) but assumes a perfect correspondence between model and reality and does not capture the deviations that occur. Documenting drift therefore requires novel computational representations to capture and share the gaps between the assumed digital model and the physical as-built state.

3DGS offers an alternative representation for documenting the as-built state of a construction project (Kerbl et al. 2023). 3DGS has been previously extended to contexts requiring in-situ capture such as machine maintenance collaboration (Kwatra et al. 2025) and architectural heritage preservation (Yu et al. 2025). Capturing improvisations requires a video recording using any camera (Batra et al. 2025) without needing site connectivity (critical for remote construction sites like forests). Existing 3DGS workflows explored a spatially grounded reconstruction that can be aligned with the original 3D model (Kwatra et al. 2025). We extend this method of geometric alignment with Pentimento, integrating on-site voice-recorded adaptations spatially, temporally, and semantically with the model-aligned splats of construction states.

## METHOD

### ReShelter Case Study

To preliminarily explore how building drift can be made legible to stakeholders, we conducted a case study of ReShelter, a shelter designed to investigate the use of reclaimed timber across multiple resource streams. In ReShelter, complex double curved glulam members from previously

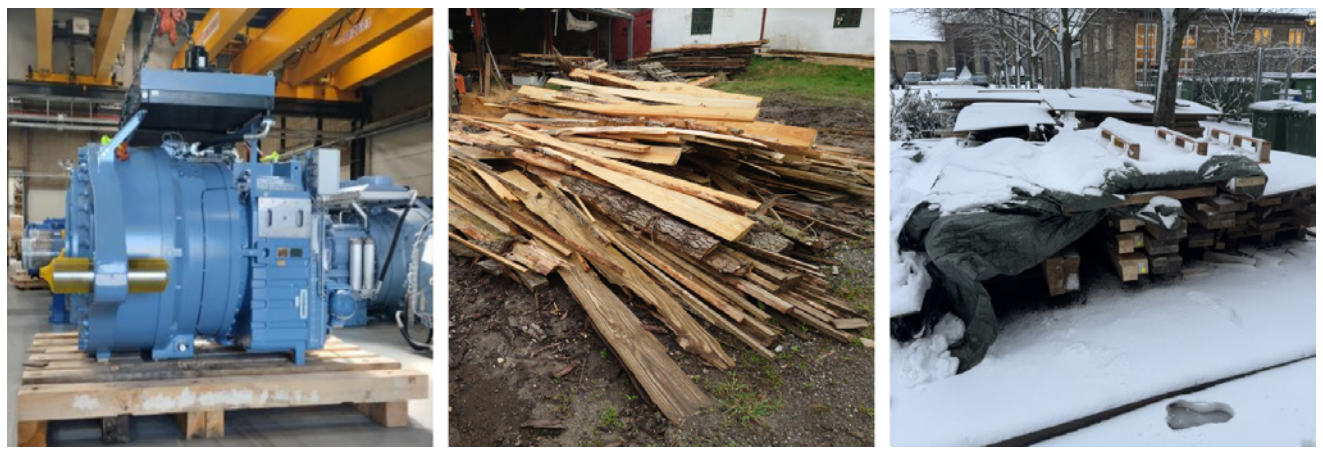

3 Examples of reclaimed material sources and storage for ReShelter.

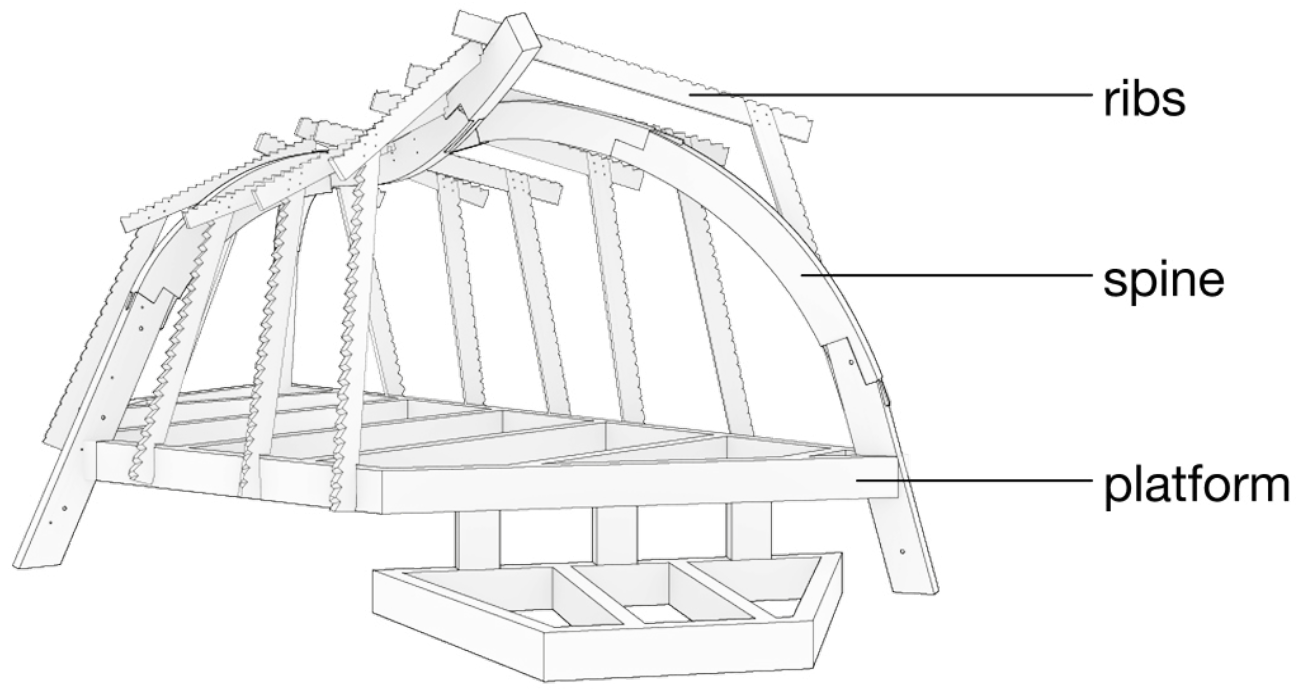

**4** ReShelter design and components.

deployed timber-related research projects (Tamke et al. 2021) combined with solid timber profiles from wind turbine transport crates and bark offcuts from a local sawmill (Figure 3). Prior to site assembly, the pieces were fabricated to closely match the designed model (Figure 4) using a robotic mill. To document the ReShelter construction process, the research team recorded videos using mobile devices and Ray-Ban Meta AI glasses, participated in the construction project as builders, and engaged with stakeholders to understand their information needs across material lifecycles. We drew on principles of qualitative research (Glaser 1968) to analyze extensive field notes from this case study conducted from January to May 2026.

ReShelter's construction proceeded in two phases as shown in Figure 5. The structure was first assembled as a dry fit in an auditorium over three days, which allowed the team to identify challenges early and start using Pentimento. The structure was then disassembled and reassembled in the forest site for final installation, where the unpredictability of the environment introduced additional sources of drift missing in the indoor assembly. Both phases informed our classification and the design of Pentimento around real-world construction environments.

### Building Drift Classification

**Tending the Site** describes drift arising from the environmental conditions of the construction site and the preparation and negotiation work that precedes and underlies the structure's visible features. A construction site is rarely a blank canvas; it has its own conditions with roots, slopes, weather patterns, and vegetation that must be worked with rather than simply built upon. This hidden labor of clearing vegetation, leveling ground, bracing against wind, and preparing foundations shapes the as-built state in ways the model cannot anticipate. For ReShelter in the forest, an unexpected tree root required breaking a concrete piece of the spine foot and repositioning it around the root, which became structurally integrated into the foot (Figure 6).

**Foraging for Fit** describes drift caused by working within the constraints of available materials and tools on site. Unlike urban construction with ready access to suppliers, remote settings require working with what is on hand: a beam too thick must be trimmed, a stone too tall must be shimmed, and a precise cut is approximated with a jigsaw. For ReShelter, the bark cladding was sourced from a nearby sawmill, but pieces varied significantly in thickness and flexibility, requiring careful selection and trimming to match the design (Figure 7).

**Interpreting the Material** describes drift arising from the inherent behavior and hidden histories of reclaimed materials. Unlike standardized materials that arrive with known, predictable properties, reclaimed materials carry tendencies that are often not visible at first encounter. They require different kinds of sensemaking, such as physically examining a timber beam to find where moisture has softened it, drilling in and discovering a hidden nail, or interpreting incoherent markings left by prior owners. For ReShelter, the glulam spine pieces required heavy knocking to fit due to moisture-driven expansion, revealing histories only apparent during construction (Figure 8).

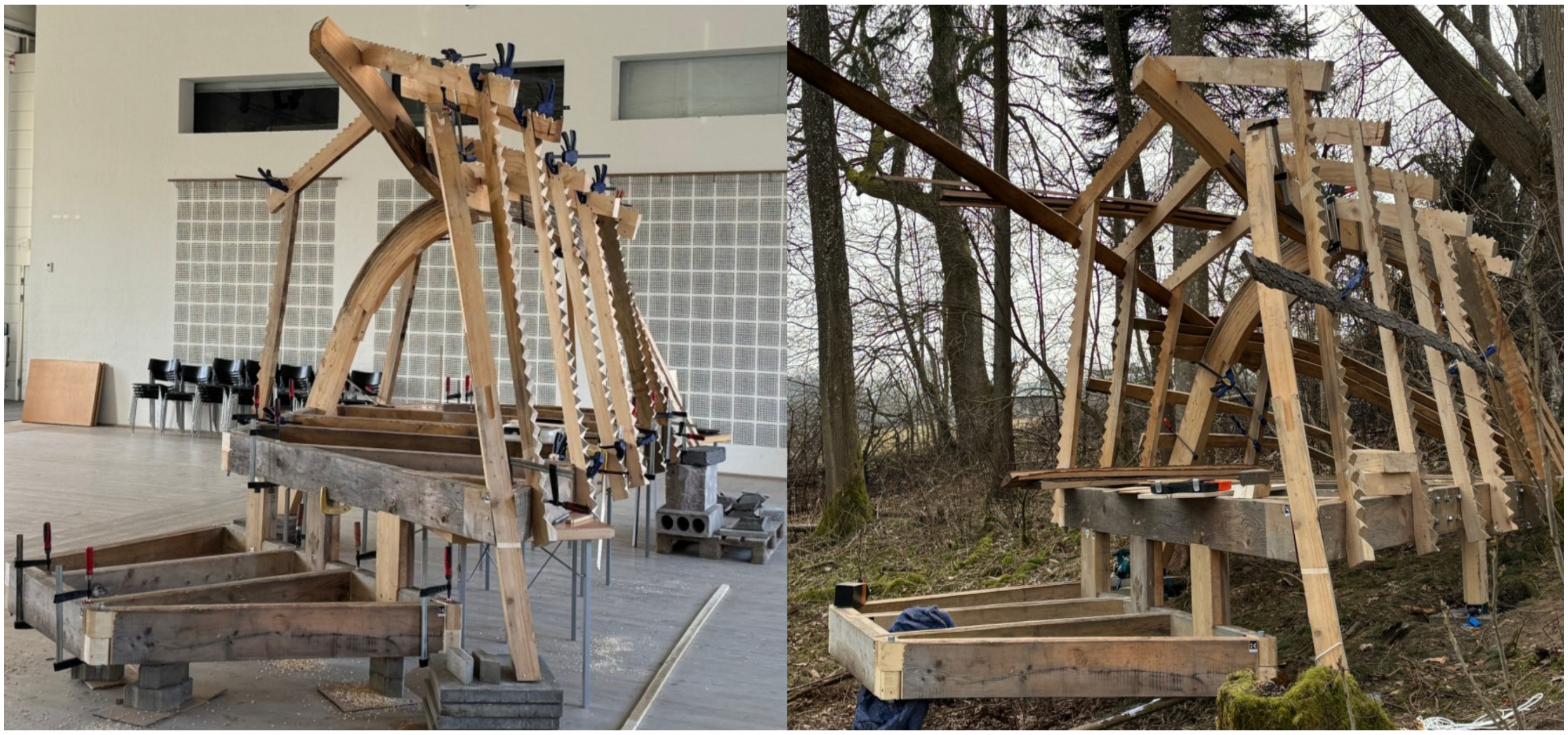

**5** ReShelter construction phases.

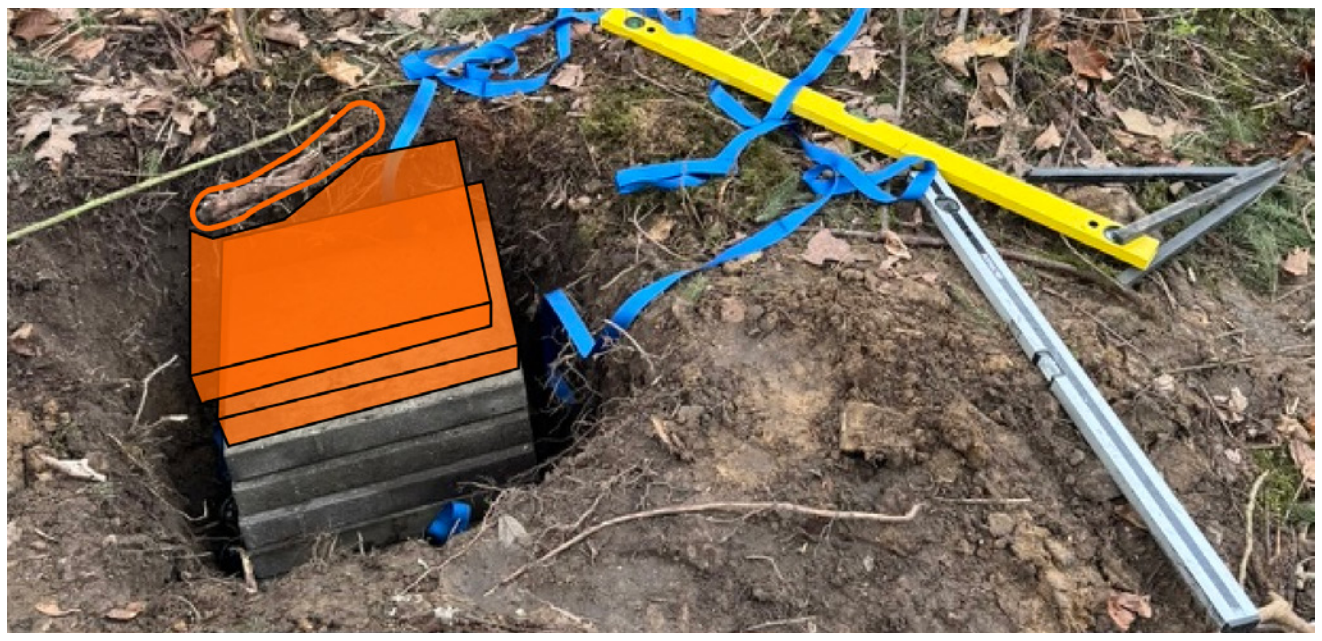

6 Tending the Site from working around a tree root.

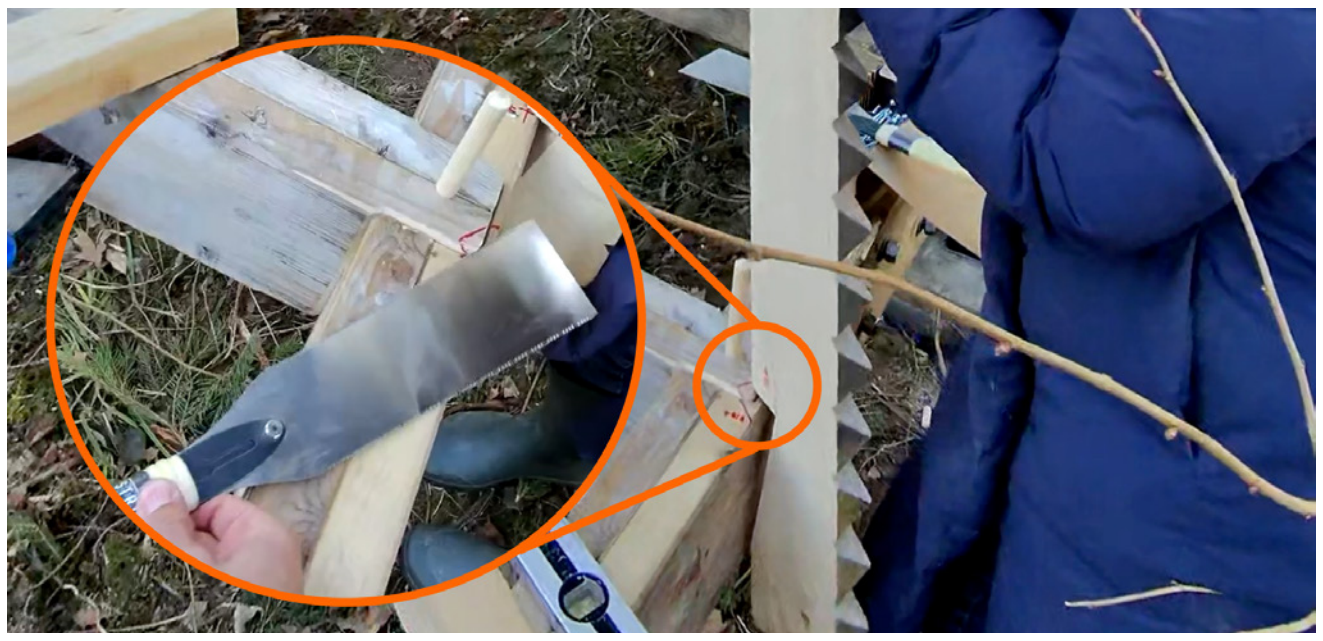

9 Marking Measurements and refining with portable tools.

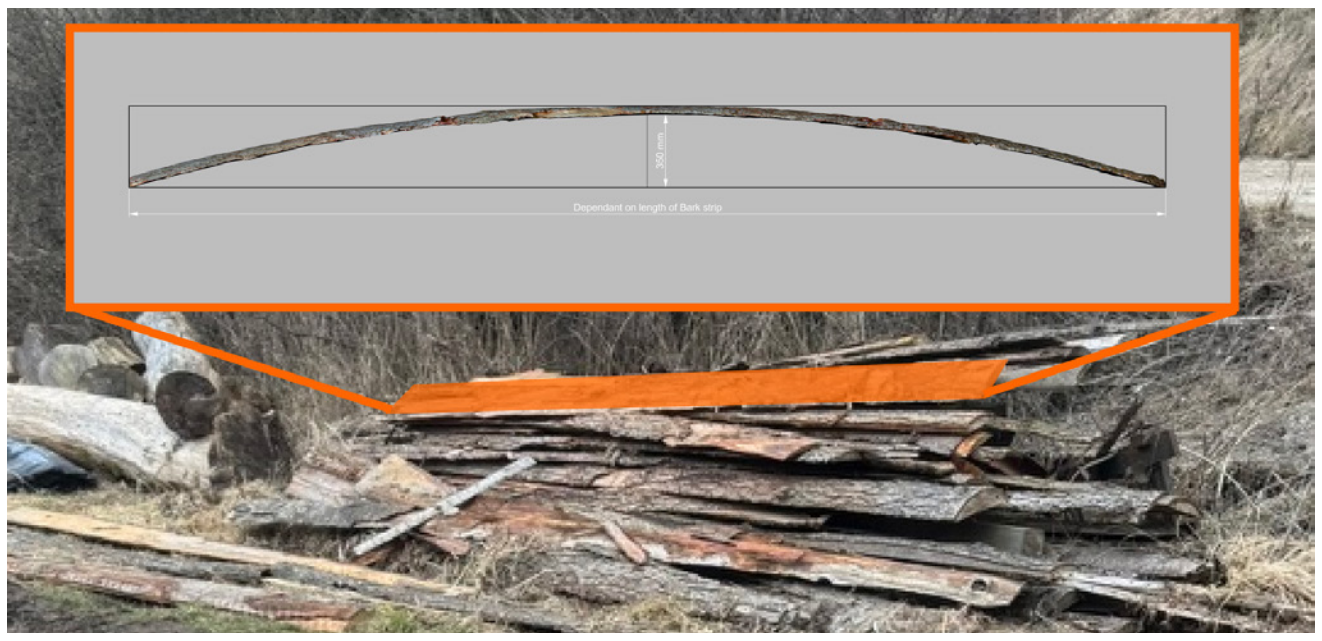

7 Foraging for Fit with various bark conditions.

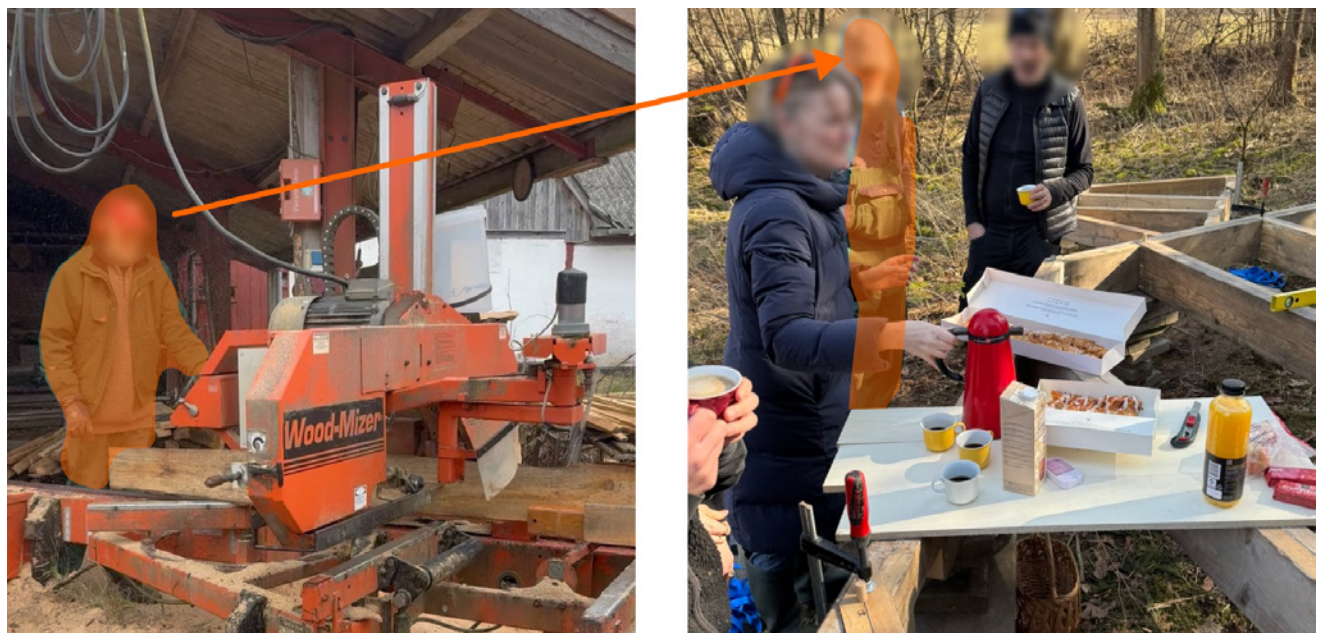

10 Coordinating Across Communities and engaging with them in different ways.

**Marking Measurements** describes drift arising from the natural imprecision of tool-constrained construction and the gap between fabrication-grade and on-site tools. Every pencil mark and hand tool cut carries imprecision that compounds across a structure, and the physical demands of balancing, reaching, and holding on site make precision challenging. ReShelter's components were initially fabricated with a robotic mill capable of millimeter precision, but that precision could not be replicated on site with portable hand tools. Rough cuts made with a portable jigsaw required further refinement with a Japanese saw, and penciled markings from the indoor dry fit indicating rib locations were difficult to interpret when moving to the forest (Figure 9).

**Coordinating Across Communities** describes drift arising from navigating the different knowledge, roles, and expectations of the communities surrounding a construction project. Construction is rarely an isolated act; site owners, designers, builders, and inhabitants each shape what gets built and where, often in ways that emerge once construction begins. These social negotiations are as much a part of the as-built state as the material and environment conditions yet are incompatible with existing digital models. Three distinct groups shaped ReShelter's construction: the core design team with full knowledge of the design intent, lab members coming to help with partial context, and the landowners with their own relationship to the site and its use. Each group introduced adaptations when expectations diverged on-site. For example, the site landowners specified after fabrication had completed that the previously selected location was too visible from the street, requiring the entire structure to be repositioned to a new area of the forest with different slope and tree locations. This negotiation evolved into an ongoing collaboration, as the landowners also shared offcuts from their local sawmill and frequently engaged with the team during construction (Figure 10).

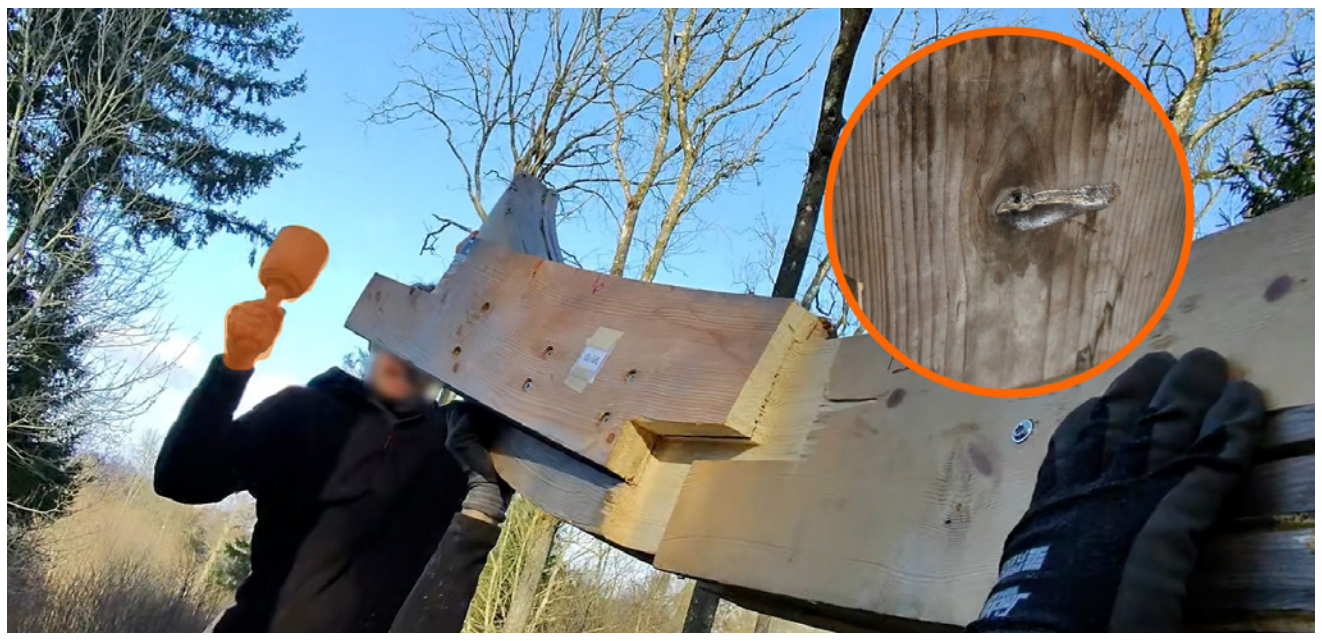

8 Interpreting the Material and forcing two moisture-expanded beams together.

### Pentimento Implementation

Pentimento is designed around four considerations: be accessible to any stakeholder through a web browser; require no site infrastructure for capture; preserve all data but can be filtered based on our drift classification; and have processing times under ten minutes per video given the volume of recordings.

With these design considerations, we developed Pentimento as a web interface with a backend processing pipeline (Figure 11). Builders capture the current state and on-site adaptations by recording video and audio on a smartphone or Meta Ray-Ban glasses for hands-free recording while narrating what they did and why for each adaptation. Whenever they have internet connectivity, builders can upload their recordings to a remote GPU instance (Nvidia 4080S, 24GB VRAM) where our system extracts frames at 4fps to balance reconstruction quality with processing time, runs a COLMAP structure-from-motion (SfM) script to estimate camera poses (Schonberger and Frahm 2016), and generates a 3DGS reconstruction using the Python package, gsplat (Ye et al. 2025). 3DGS represents a captured scene as a cloud of semi-transparent, colored Gaussians whose positions, sizes, and orientations are iteratively optimized by comparing rendered views against the captured video frames. The result is a photorealistic, navigable 3D scene that can be rendered from any viewpoint within a web browser and therefore shared with others.

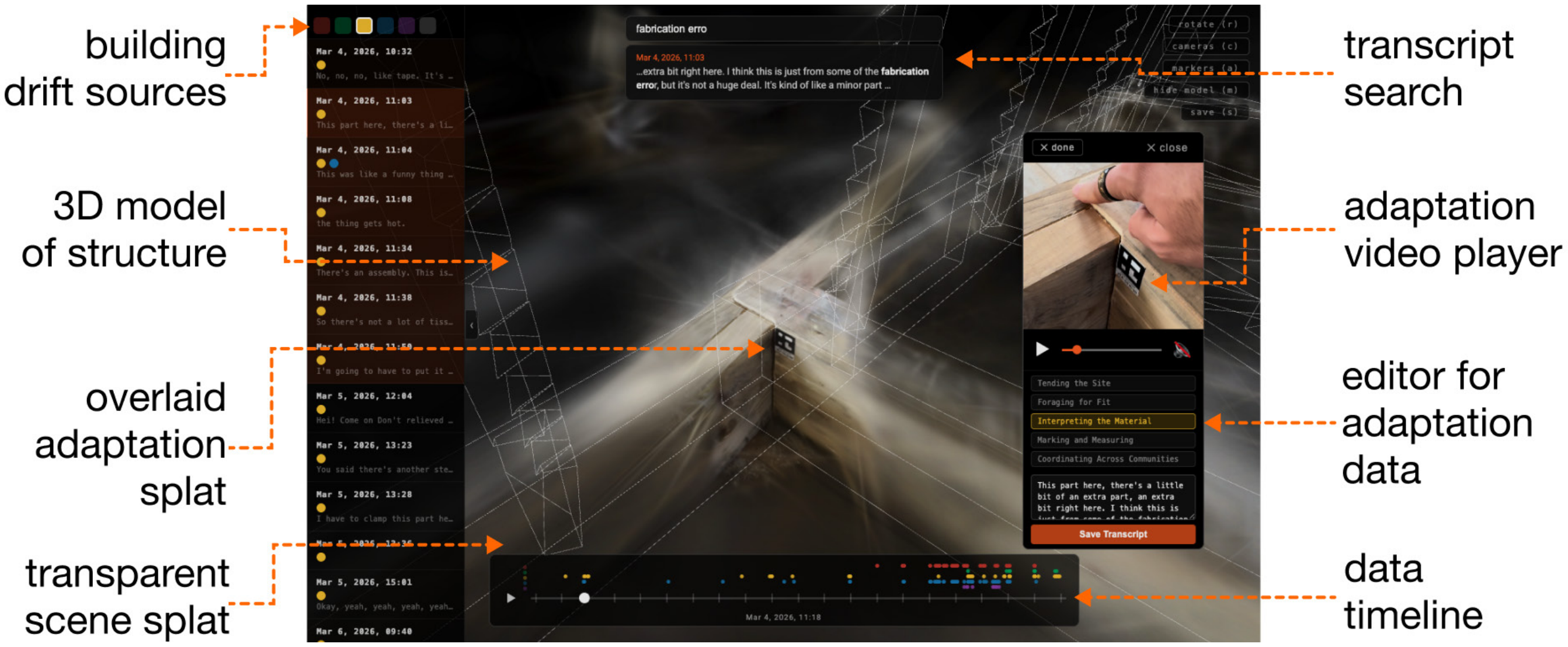

**11** Pentimento interface features.

Pentimento represents sources of building drift across three dimensions: spatially, through alignment of 3DGS with original model; temporally, through a chronological record of when and under what conditions adaptations were made; and semantically, through transcribed voice recordings categorized into the sources of building drift.

For spatial representation, we extend the SplatOverflow methodology (Kwatra et al. 2025) of placing ArUco marker stickers (Munoz-Salinas 2012) on the physical structure at locations corresponding to marked squares in a glTF representation of the model (exported by CAD tools such as Rhino). The backend detects these markers and calculates a weighted transform (rotation, scale, translation) to align the splat to the model based on the markers detected and the diversity of viewing angles across frames.

For temporal representation, we use the videos' timestamps to place adaptations and scenes in chronological order, allowing stakeholders to navigate the construction process and understand how adaptations compounded. Adaptations are anchored using the same marker-based alignment but registered against the most recently captured scene so the change is most visible.

For semantic representation, voice recordings for adaptations are transcribed using a local instance of OpenAI Whisper (Radford et al. 2023). The transcripts along with the first, last, and center frame of the video are then shared with the Claude API (Anthropic 2023), prompted with descriptions of the five sources of building drift and tasked to categorize each adaptation into at least one of them. Both resulting categories and transcripts can be manually edited in the interface.

All project artifacts (CAD model, 3DGS, adaptation videos, video transcripts, calculated camera positions) are stored in a structured AWS S3 bucket (Amazon Web Services 2024). The frontend web interface pulls from the bucket to render these artifacts together with Three.js (Cabello et al. 2010) and Spark (World Labs 2025) libraries. The full workflow of Pentimento is shown in Figure 12.

## RESULTS

### Spatially

Pentimento's spatial grounding enabled stakeholders to locate and interpret sources of drift relative to the original model. In Figure 13a, a builder took a measurement of a bolt and documented it in place, anchored to the designed geometry. An inheritor reusing that timber beam would see not just the measurement but how it was measured, information invisible in both the model and a conventional site photograph. In Figure 13b, a builder described an acceptable tolerance for a joint, combining spatial reference with reasoning in a way that neither text nor geometry can provide. Video here acts as a spatial mechanism, allowing builders to gesture at specific parts of the structure in ways that written annotations cannot replicate.

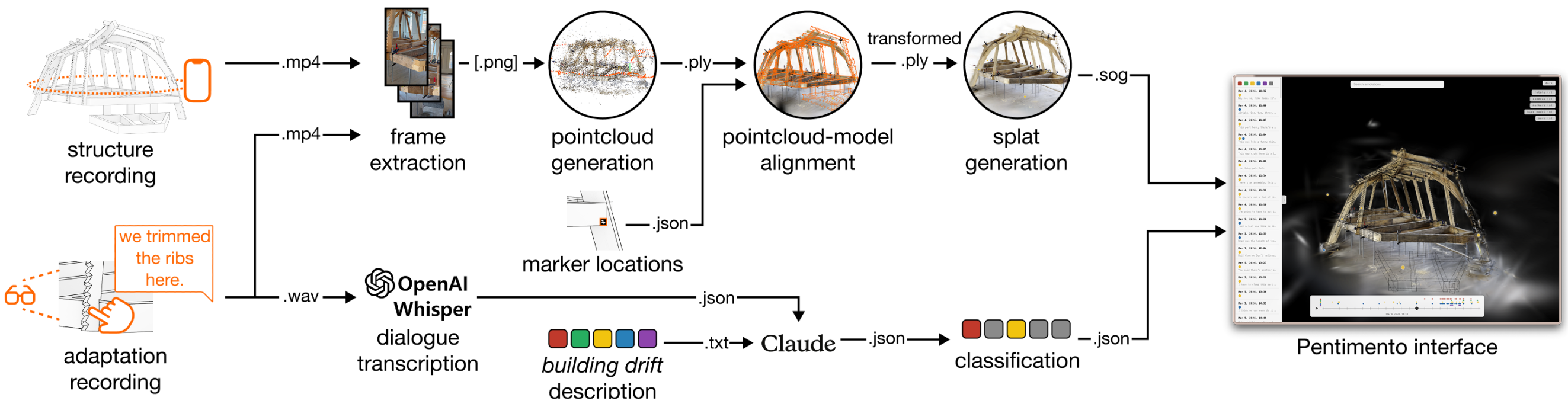

**12** Pentimento workflow.

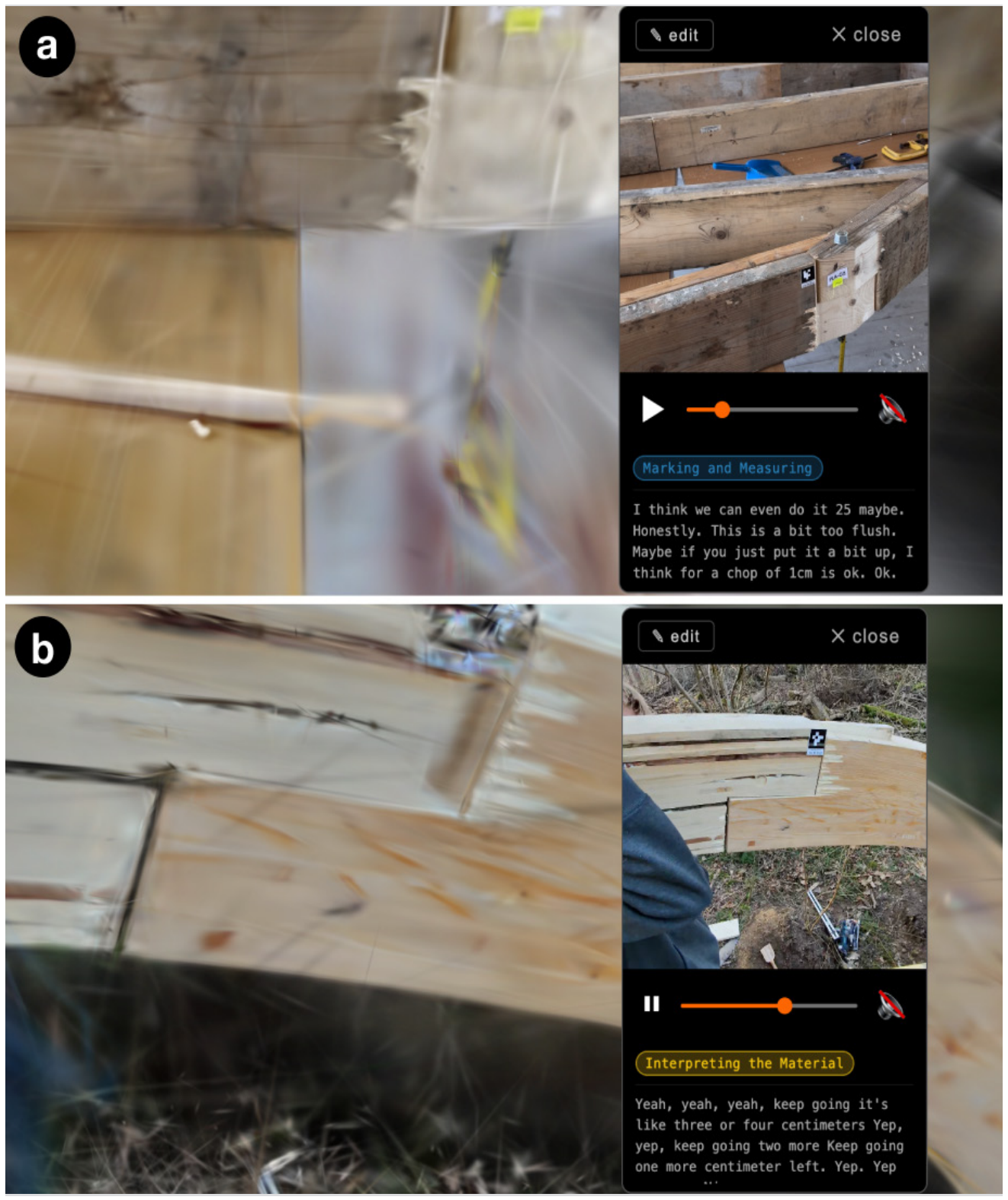

13 Examples of overlaid adaptation splats in Pentimento.

## Temporally

Pentimento's temporal record enabled stakeholders to navigate the construction process as it unfolded rather than as a single finished state. As shown in Figure 14, annotations were organized chronologically and filtered by drift source, allowing a collaborator to see what happened on a given day and in what order. Referencing the sequence of assembly was crucial for understanding why deviations occurred and how they

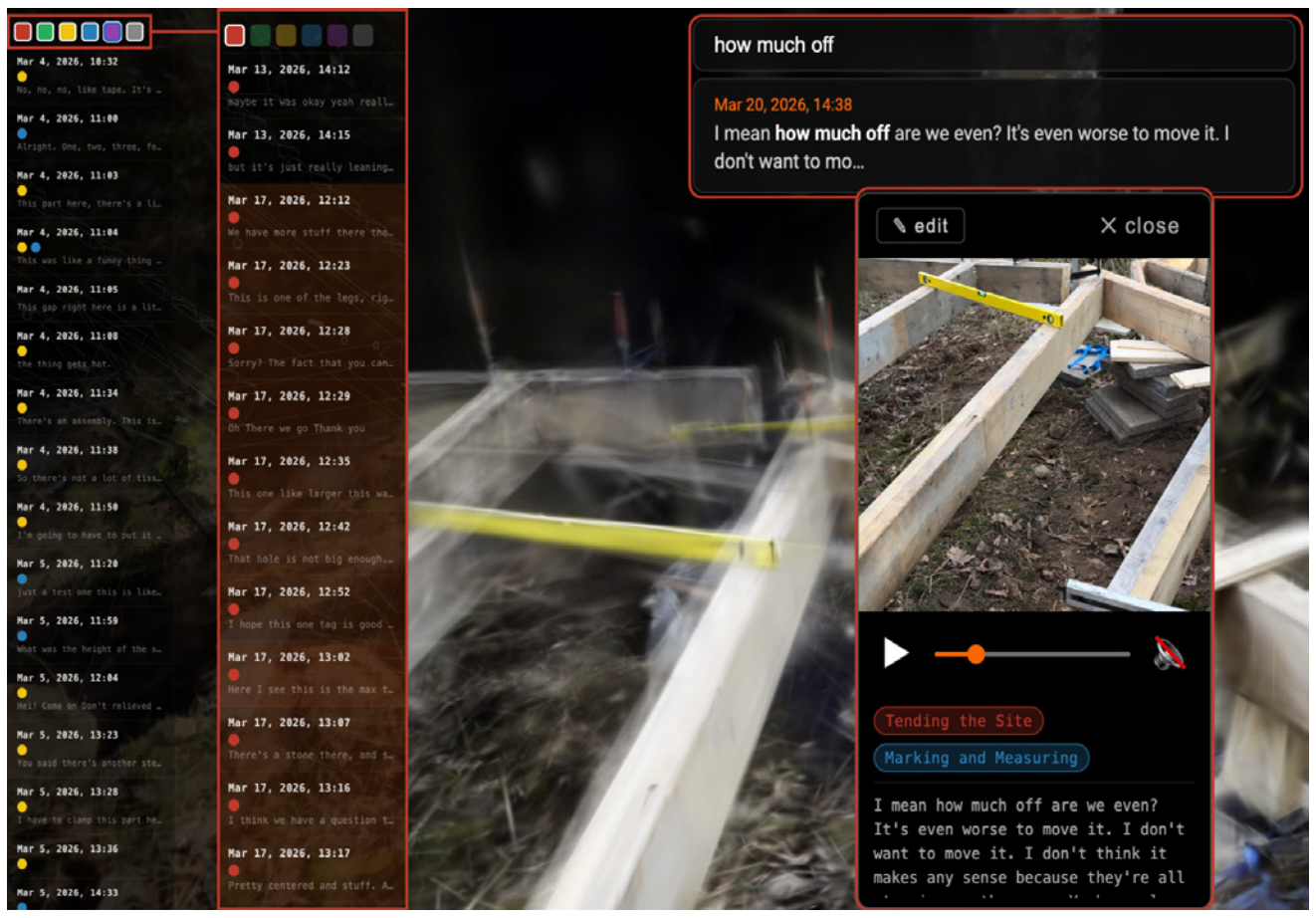

15 Search and filter features of Pentimento.

compounded. For example, the team discovered during the dry run that the spine setup sequence needed reversing, informing the forest installation. Environmental conditions also become visible temporally: a rainy-day during forest installation affected which components could be safely handled, and this context is recoverable through the dated annotation record. An evaluator assessing a material's reuse potential can therefore understand not just what was built but under what conditions, distinguishing weather-driven drift from deliberate design decisions.

## Semantically

Lastly, Pentimento's semantic record captures the reasoning behind adaptations that spatial reconstruction cannot convey. In Figure 15, the transcript preserves a tolerance judgement: *"How much off are we even? It's even worse to move it."* A collaborator or evaluator can recover not just what was decided but why, distinguishing an intentional acceptance from an overlooked error. Crucially, the inheritor can also use this record to confidently dismiss a deviation rather than treating it as an unknown risk.

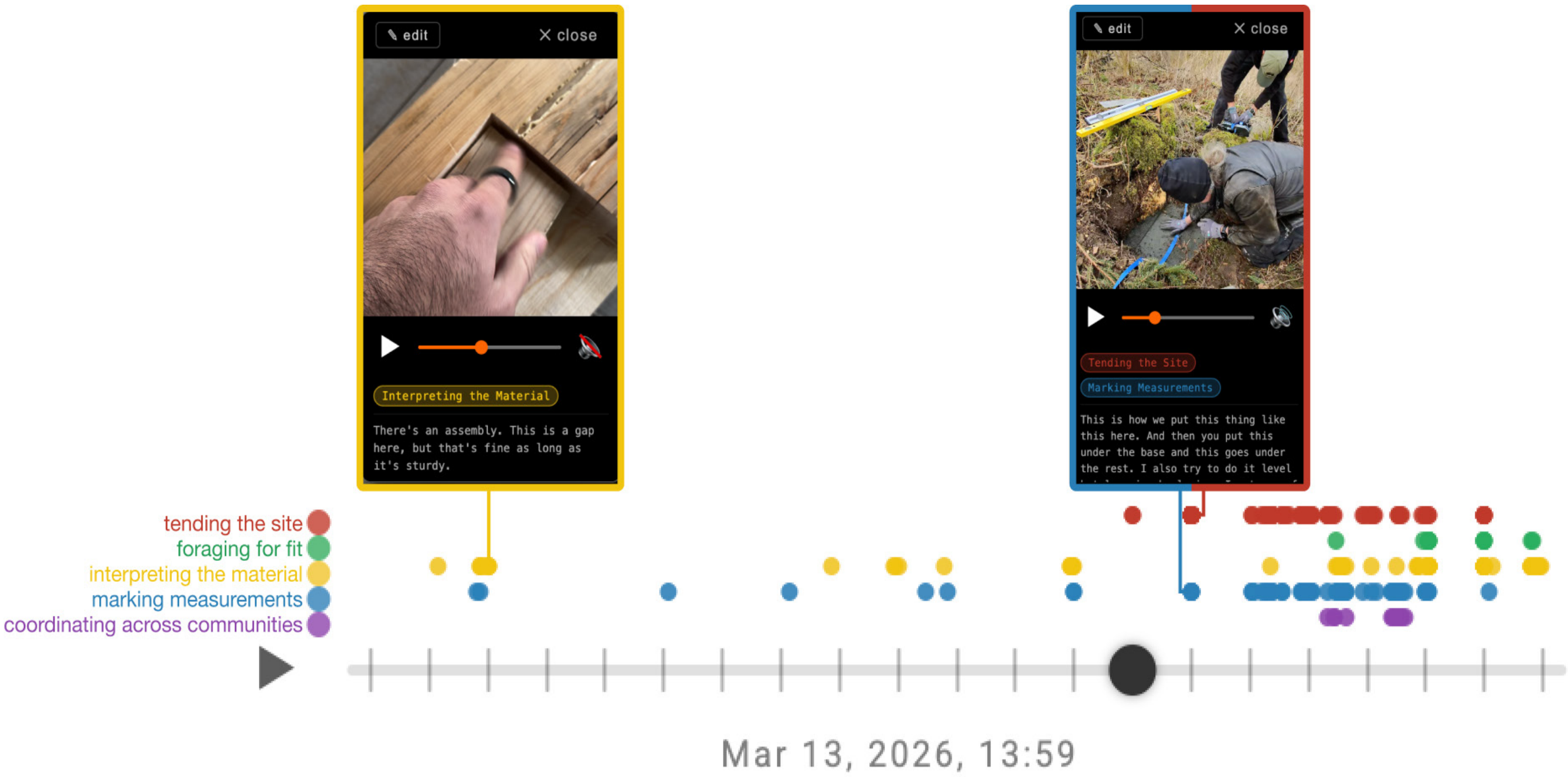

14 Pentimento timeline with example sources of drift.

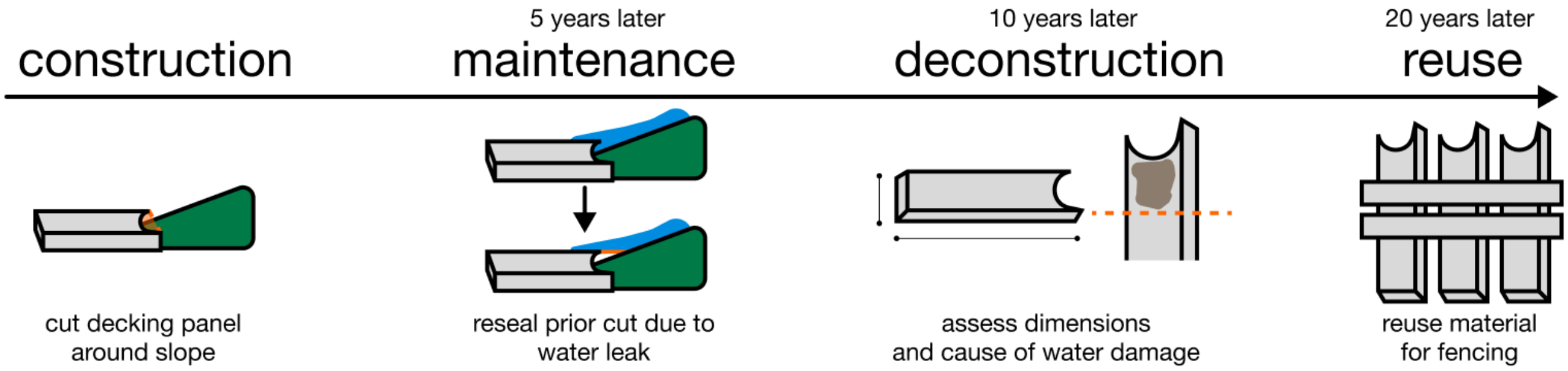

**16** Different use cases for Pentimento across the lifecycle of structures.

The ability to tag adaptations with multiple drift sources further reflects the multidimensional nature of building drift: one adaptation does not necessarily belong to one category alone and forcing it into one would produce an incomplete and partial record. The search and filter features shown in Figure 15 allow each stakeholder to navigate the annotations relevant to them, making the record legible via selectivity.

## DISCUSSION

### Documenting on Site

Deploying Pentimento for ReShelter revealed that the same heterogeneity that makes building drift inevitable also makes it difficult to document. ArUco markers often fell off timber surfaces due to wood shavings and weather, requiring constant reattachment. Capturing complete splats requires a steady loop around the structure, but navigating forest shrubs, uneven terrain, and snow led to shaky videos and blurry splats. Participating as a builder also made timely documentation difficult without interrupting the work. These challenges reflect a broader tension between the improvisational and messy nature of construction sites and the stable, pristine conditions that computer vision tools typically assume (Teizer 2015). Portable cameras offer a flexible pathway to document improvisational working environments without modifying or intruding on the site, even if the resulting captures are imperfect.

### Building Layered Digital Models

Pentimento's response to imperfect data capture points towards a different kind of digital model for architecture. While standard CAD tools embody Ingold's hylomorphic model of making with the outcomes of construction, Pentimento offers a layered architecture for materials and environments to "push back" against the designed model by preserving how builders responded to that resistance (Ingold 2009). The reasoning behind these responses in aggregate is what stakeholders need to effectively reuse materials across multiple building lifetimes. Pentimento maintains multiple partial splats, videos, transcripts, and CAD geometries that each stakeholder can draw on depending on what they need, together capturing more than any single representation could. As shown by the speculative examples in Figure 16, a maintainer investigating a leak could filter Pentimento for building elements in the affected area and see a decking panel had been recut on site to redirect runoff around an unexpected slope, while a deconstruction crew could later consult the same record to confirm the panel's true dimensions and the maintainer's repairs before reusing it elsewhere. These layers carry a material's history across building lifetimes, preserving not just what it became but how and why.

Our system and classification for sources of building drift are partial and preliminary, drawn from one case study, but offer an approach that can be further tested and extended across other reclaimed-material building projects. Our project relies on the qualitative assessment of splat generation and alignment, but future work may explore quantifying drift and splat precision through dedicated 3DGS metrics. And just as materials persist across building lifetimes, it is critical to explore alternative methods of maintaining the digital record. Saving annotations, transcripts, and videos as universal and parseable formats such as JSON is one way to ensure that this record remains accessible even as tools and technologies advance.

### Designing for Drift

Embracing building drift could inform future computational workflows by guiding designers on how drift may emerge during fabrication and on-site construction. For example, on-site tablet viewing of material histories could enable collaborators to see the aligned model overlaid on the structure as they work. Deploying drift-inspired tools at scale also requires research into how multiple contributors can record simultaneously and how overlapping or contradictory records can be reconciled. Finally, recent advances in AI could draw on this knowledge to personalize capture, analysis, and consolidation to each stakeholder and identify adaptation patterns across projects.

Integrating drift into CAD tools has broader implications for sustainable construction. While architecture presumes control through an authoritative model, building drift reveals that buildings are not entirely as planned, materials remain dynamic, and sustainable practices require thinking further into the pre-lives and post-lives of materials. Materials have micro ecosystems, environments change through the seasons, and local communities shift and flow; all these considerations require design tools that allow imperfections and adaptations to emerge in conversation with the world, rather than treating them as failures to be corrected. A layered, imperfect digital model such as Pentimento offers a record of these conversations in support of a circular economy in which materials travel across building lifetimes, human actors come and go, and the vagaries of weather, chance, and environment reshape both wholes and parts.

## CONCLUSION

Building drift is inevitable when working at a site, with a material, and across the communities surrounding a construction project. In this paper, we introduce building drift as a critical concept for sustainability and characterize five of its sources through the construction of ReShelter in a forest. We demonstrate how to document these sources of drift through the development of Pentimento. By spatially, temporally, and semantically representing on-site adaptations in relation to the original digital model, this work begins to close the gap between what was designed and what was built. Our vision for this work is to inspire construction and, more broadly, digital fabrication practices that do not merely tolerate drift in individual projects but use drift as a mechanism to connect them through the preserved history of adaptations made to each material and go beyond geometry to preserve the social and ecological dimensions of how each material came to be what it is. For a truly circular economy, that history supports making material repair, recovery, and reuse practical for construction projects and meaningfully connected to the people and places that shaped them.

## ACKNOWLEDGMENTS

We thank Carolyn-Nelle Preston Ichniowski, Lena Fritsch, Nadja Gaudillière-Jami, Katerina Toumpektsi, and Amin Adelzadeh for their insights during this project and Torben for sharing his wisdom, forest space, coffee, and offcuts with us. The ReShelter project has received funding from the European Research Council (ERC) under the European Union's Horizon 2020 research and innovation programme (grant agreement No 101019693).

## IMAGE CREDITS

All drawings and images by the authors.

**Ritik Batra** is an Information Science PhD candidate at Cornell Tech in New York City. As part of this project collaboration, he visited the Centre of Information Technology and Architecture (CITA) in Copenhagen during Spring 2026. His research focuses on computational tools for documenting and sharing expert knowledge around physical and digital materials. He has industry experience at Adobe, Autodesk, Stripe, Asana, and Goldman Sachs. Before graduate school, he earned his B.S. from UC Berkeley in electrical engineering and computer sciences with minors in political economy and data science. For more information, check out his website: https://ritikbatra.com

**Martin Tamke** is Associate Professor at the Centre for Information Technology and Architecture (CITA) in Copenhagen. He pursues design-led computational strategies and technologies for transforming the building industry towards sustainable and circular practices with bio-based materials. His interdisciplinary research spans scales from harvested to living materials. Martin joined the newly founded research centre CITA in 2006 and helped shape its design-based research practice. He was appointed Guest Professor at the University of Stuttgart's Cluster of Excellence IntCDC in 2022 and is currently Guest Professor at Politecnico di Milano. He is involved in the ERC project Eco-Metabolistic Architecture (EMA) and, since 2024, has led the EIC project RAW: Computation for a New Age of Resource Aware Architecture.

**Tom Svilans** develops solutions for the digital design, construction, and manufacturing of timber projects. Tom is currently CTO at Omtre AS and a researcher at the Royal Danish Academy in Copenhagen, Denmark. His work focuses on digital design across the timber value chain - from forestry to construction and reuse - and how imaging and communication technologies can link design intelligence with timber fabrication. As a consultant, he specializes in the computational modelling, design development, and fabrication coordination of complex timber structures.

**Jan Hüls** holds a Master's degree from the Computation in Architecture programme at the Royal Danish Academy and is currently a Research Assistant at the Centre for Information Technology and Architecture (CITA). Previously, he worked as a Student Research Assistant at the Institute for Structural Design (ITE) at the Technical University of Braunschweig, contributing to research in digital fabrication and additive manufacturing. His work explores computational design strategies, robotic fabrication, and material-driven approaches to architecture, with a current focus on digital timber fabrication and repair techniques for reclaimed materials.

**Amritansh Kwatra** is an independent researcher based in New York City. His work examines computers, machines & their maintenance. He spends a lot of time reading things on the internet.

**Steven J. Jackson** (https://sjackson.infosci.cornell.edu/) is a Professor of Information Science and Science & Technology Studies and Vice-Provost for Academic Innovation at Cornell University. He teaches and conducts research in the areas of maintenance, repair and sustainability; infrastructure, learning, and collaboration; and technology, inequality and global change. His work draws on research in Human-Computer Interaction and Computer-Supported Cooperative Work along with social science and humanities fields from STS, media studies, and anthropology to philosophy, law, and public policy. He also directs the Computing On Earth Lab, an experimental collaboration that brings together social scientists, humanists, artists and engineers to rethink the material and planetary foundations of computing.

**Thijs Roumen** is Assistant Professor in Information Science at Cornell Tech, NYC. He teaches and conducts research in the domain of Human Computer Interaction. His work focuses on computational fabrication and accessibility. He directs the Matter of Techlab since 2022 and is founding director of the Cornell Initiative on Accessible Technology and AI (CATAI). He did his PhD with Prof Patrick Baudisch at HPI, Potsdam, holds a MSc in IT Product Design from SDU, Sønderborg, and a BSc in Industrial Design from TU/e Eindhoven.

**Mette Ramsgaard Thomsen** is Professor at the Royal Danish Academy and founder of the Centre for IT and Architecture (CITA). Her research examines the intersection of architecture, computation, and material systems, with a particular focus on bio based design, digital fabrication, and sustainability. She is Principal Investigator of the ERC Advanced Grant Eco Metabolistic Architecture (EMA), which investigates regenerative approaches to architecture through computational design and bio based materials. She has led several major international research initiatives, including InnoChain, and has held the Paul Philippe Cret Chair at the University of Pennsylvania.